\documentclass[12pt,preprint]{aastex}
\usepackage{amsmath}
\usepackage{graphicx}
\usepackage{ulem}
\begin{document}

\title{Discovery of state transition behaviors in PSR J1124--5916}
\author{M. Y. Ge$^{1}$, J. P. Yuan$^{2,3}$, F. J. Lu$^{1}$, H. Tong$^{4}$, S. Q. Zhou$^{5}$, L. L. Yan$^{6}$, L. J. Wang$^{1}$, Y. L.Tuo$^{1,7}$, X. F. Li$^{1}$, L. M. Song$^{1,7}$}

\affil{$^{1}$ Key Laboratory of Particle Astrophysics, Institute of High Energy Physics, 
Chinese Academy of Sciences, Beijing 100049, China. Email: gemy@ihep.ac.cn}
\affil{$^{2}$ Xinjiang Astronomical Observatory, Chinese Academy of Sciences, Xinjiang 830011, China}
\affil{$^{3}$ Center for Astronomical Mega-Science, Chinese Academy of Sciences, Beijing, 100012, China}
\affil{$^{4}$ School of Physics and Electronic Engineering, Guangzhou University, Guangzhou 510006, China}
\affil{$^{5}$ China West Normal University, Sichuang 637002, China}
\affil{$^{6}$ School of Mathematics and Physics, Anhui Jianzhu University, Hefei 230601, China}
\affil{$^{7}$ University of Chinese Academy of Sciences, Chinese Academy of Sciences, Beijing 100049, China}
\begin{abstract}

With the twelve-year long observations by {\sl Fermi}-LAT, we discover two pairs of spin-down state transitions 
of PSR J1124--5916, making it the second young pulsar detected to have such behaviors. 
PSR J1124--5916 shows mainly two states according to its spin-down rate evolution, the normal 
spin-down state and the low spin-down state. In about 80\% of the observation time, the 
pulsar is in the normal spin-down state, in which the spin-down rate
decreases linearly and gives a braking index of $1.98\pm0.04$. 
The two transitions to the low spin-down state are 
 in MJD 55183--55803 and MJD 56114--56398 respectively, with fractional amplitudes both $\sim0.4\%$.
No significant difference between the $\gamma$-ray profiles of the two spin-down states is 
detected, which is similar to PSR B0540-69, the other young pulsar with state transition detected.

\end{abstract}


\section{Introduction}

The rotation evolution of a pulsar reflects the properties of its magnetosphere.
For pulsars, the rotation frequency $\nu$ decreases with time and this slow-down is 
usually described by the relation $\dot{\nu}=-K\nu^{n}$ (\cite{2011ApJ...741L..13E}, and references therein), 
where $K$ is a positive constant depending on the moment of inertia and the magnetic dipole
moment of the neutron star, and $n$ is the braking index derived with
 $n={\nu\ddot{\nu}}/{\dot{\nu}^{2}}$, which is related to the physics mechanism slowing down the 
pulsar rotation, and $n=3$ is expected if the
spin-down is dominated by pure magnetodipole radiation \citep{1977puls.book.....M}. Currently,
only twelve pulsars have reliably measured braking indices because the timing 
noise and stochastic variations of the pulsar spin-down
make accurate determinations of the braking index very difficult. The measured $n$ values are
in the range 0.9--3.15 \citep{2011ApJ...741L..13E,2017MNRAS.466..147E,Archibald2016,Clark2016},

Some pulsars show rotation state transition behaviors.
The emission of PSR B1931+24 switches on and off periodically, and the transition 
could finish in 10\,seconds \citep{2006Sci...312..549K}. \cite{2010Sci...329..408L} reported that some
pulsars show correlation between the variation of spin-down rate $|\dot{\nu}|$
and pulse profile change. PSR J2021+4026 is the 
first variable $\gamma$-ray pulsar, whose $\gamma$-ray flux and pulse profile change 
with $|\dot{\nu}|$ variations \citep{2013ApJ...777L...2A,2017ApJ...842...53Z,2020ApJ...890...16T}.
PSR B0540--69 is the first young pulsar detected to have spin-down rate transition. 
It results in a delayed brightening of its pulsar wind nebula \citep{Ge2019}.
However, the pulsed X-ray flux and profile of PSR B0540--69 show no variations in this process 
different from the previous reported pulsars. It is believed that these 
state transition behaviors are probably induced by the changes of the magnetosphere,
though the detailed mechanism is unclear.

PSR J1124--5916 is associated with supernova remnant G292.0+1.8 \citep{2002ApJ...567L..71C,2003ApJ...591L.139H}. 
It has a characteristic age ($\tau_{c}$ ) of 2900\,yr and a 
spin-down luminosity $\dot{\rm E}$ of $1.2\times10^{37}\,{\rm erg\,s^{-1}}$, making it
the sixth youngest and the sixth most energetic pulsar detected so far.
The pulse signals from this object have been detected in the radio, X-ray and $\gamma$-ray bands
\citep{2002ApJ...567L..71C,2003ApJ...591L.139H,2011ApJS..194...17R}.
With the first two years of data collected by the Large Area Telescope (LAT) of 
the {\sl Fermi} Gamma-ray Space Telescope ({\sl Fermi}), the braking index of PSR J1124--5916
is determined as $-3.78$  \citep{2011ApJS..194...17R}.
Here we report the discovery of state transitions of this pulsar and the new measurement 
of its braking index, using the twelve-year long {\sl Fermi}-LAT observations
and the previous results from \cite{2002ApJ...567L..71C}, \cite{2003ApJ...591L.139H} 
and \cite{Abdo2010b}.

\section{Observations and Timing Analysis}

LAT is the main instrument of {\sl Fermi}, which could obtain the directional
measurement, energy measurement for $\gamma$-rays, and background
discrimination, respectively \citep{Atwood2009}. The energy range of LAT 
is from 20\,MeV to 300\,GeV and the effective area is $\sim 8000$\,cm$^2$. 
As {\sl Fermi} can survey the sky frequently with also relatively long exposure, 
it is suitable to monitor the spin evolution of $\gamma$-ray pulsars.

The {\sl Fermi}-LAT observation of PSR J1124--5916 spans from MJD 54763 to 58909.
We analyze the data using the standard Fermi Science Tools (v10r0p5).
The events are firstly selected using {\it gtselect} with angular distance less than 0.5\textordmasculine, zenith angle less than 105\textordmasculine\,\, and
energy range 0.1 to 10\,GeV \citep{Abdo2010}
\footnote{https://fermi.gsfc.nasa.gov/ssc/data/analysis/scitools/pulsar\_analysis\_tutorial.html}. 
Then, the arrival time of every event is corrected to Solar 
System Barycentre (SSB) using {\it gtbary} with the solar system ephemerides DE405
and the pulsar position of $\alpha=11^{\rm h}24^{\rm m}39^{\rm s}$
and $\delta=-59^{\textordmasculine}16^{\prime}19^{\prime\prime}$ \citep {2011ApJS..194...17R}.
Each TOA is accumulated from every 15-day exposure. The detailed procedure
could be found in \cite{Ge2019}.

In order to show the spin evolution directly, we divide the data set in the whole time 
range into subsets. The time steps for $\nu$ 
and $\dot{\nu}$ analyses are 60\,days and 100\,days, respectively. 
For each subset, the coherent timing analysis is performed using TEMPO2, 
and the center of the time span is taken as the reference epoch of the 
timing analysis \citep{Hobbs et al. (2006)}. Then, the spin evolution of the pulsar in a time interval
could be fitted by equation (\ref{eq00}).
\begin{equation}
\Phi=\Phi_{0} + \nu{(t-t_{0})}+\frac{1}{2}\dot\nu{(t-t_{0})^{2}}+\frac{1}{6}\ddot\nu{(t-t_{0})^{3}} ,
\label{eq00}
\end{equation}
where $\nu$, $\dot\nu$ and $\ddot\nu$ are the spin parameters at epoch $t_{0}$.
To calculate braking index, equation (\ref{eq11}) is utilized to fit the spin evolution 
covering different time intervals for one state.
\begin{equation}
\nu=\nu_{0}+\dot\nu{(t-t_{0})}+\frac{1}{2}\ddot\nu{(t-t_{0})^{2}+\Delta\nu_{\rm J1}+\Delta\nu_{\rm J2}} ,
\label{eq11}
\end{equation}
where $\nu_{0}$, $\dot\nu$ and $\ddot\nu$ are the spin parameters at epoch $t_{0}$. $\Delta\nu_{\rm J1}$ and $\Delta\nu_{\rm J2}$ are the step values of the spin frequency due to different states.

The timing analysis results obtained by \cite{2002ApJ...567L..71C}, \cite{2003ApJ...591L.139H} 
and \cite{Abdo2010b} are included in this analysis to show the long term evolution of PSR J1124--5916.

\section{Results}

\begin{figure}
\begin{center}
\includegraphics[scale=0.75]{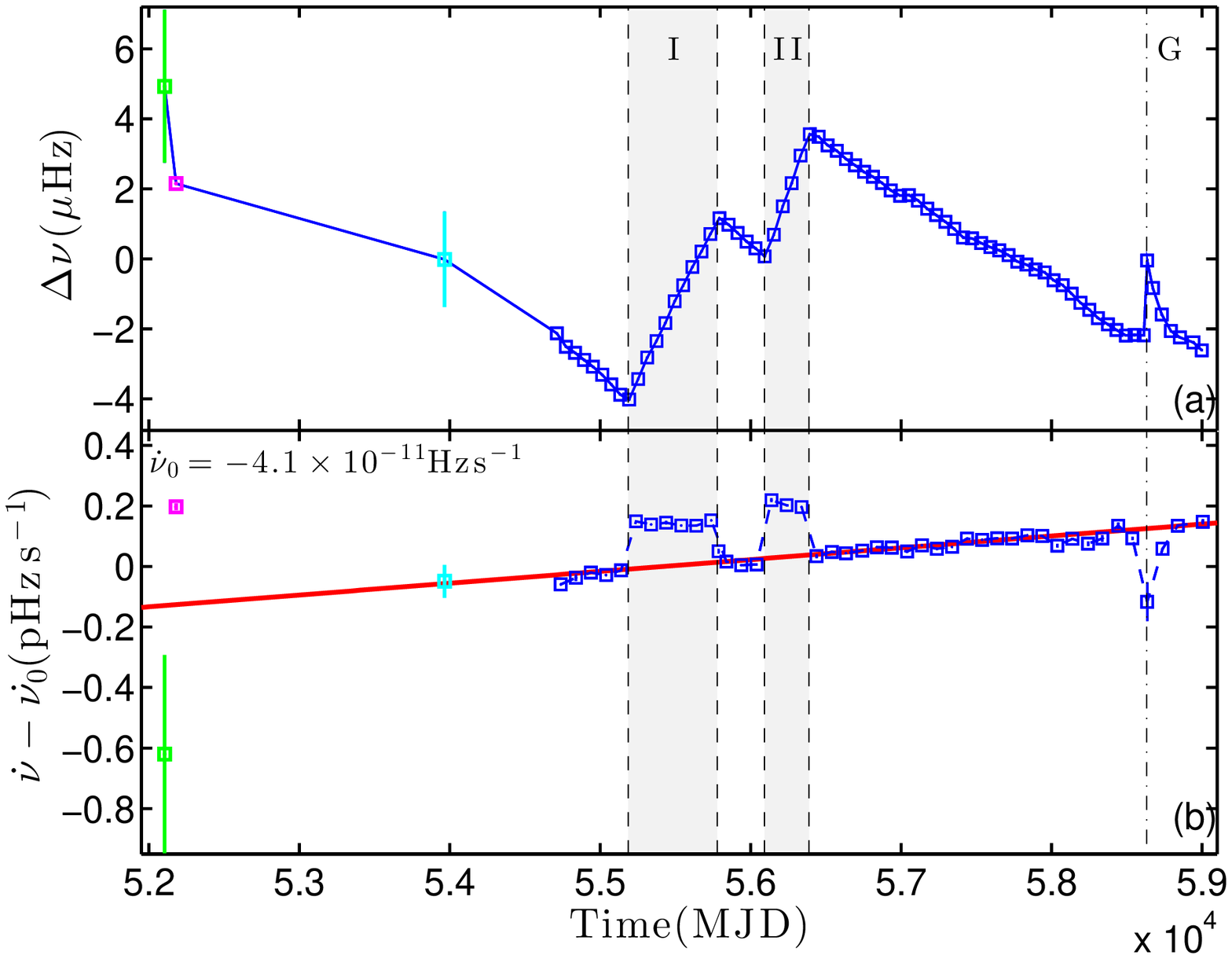}\caption{
The spin frequency evolution of PSR J1124--5916. Panel (a): The spin evolution with the 
quadratic polynomial fitting result listed in Table \ref{table_timing_para0} subtracted. The green, purple and cyan
square points are taken from \cite{2003ApJ...591L.139H}, \cite{2002ApJ...567L..71C} 
and \cite{Abdo2010b}, respectively. Panel (b): The $\dot\nu$ evolution, where $\dot\nu_{0}=-4.1\times10^{-11}\,{\rm  Hz\,s^{-1}}$. The red line represents the evolution of $\dot\nu$ obtained by fitting to the $\nu$ values of whole normal spin-down state with equation (\ref{eq11}), and the fitting parameters are listed in 
Table \ref{table_timing_para1}. The value of $\dot{\nu}$ at MJD 53965 is derived from the frequency 
at this epoch and the first frequency result on MJD 54738 by {\sl Fermi}-LAT. The four vertical dashed lines 
represent the start time and stop time of state transitions. The two gray belts 
represent the low spin-down states. The glitch epoch is marked by the dot-dashed line.
\label{fig0}}
\end{center}
\end{figure}

The spin evolution between MJD 54600 and 57000 
shows different spin-down behaviors after subtracting the long-term evolution trend represented
by the quadratic polynomial fitting, as shown in Figure \ref{fig0} (a) and listed in Table \ref{table_timing_para0}.
The shape of $\Delta{\nu}$ is like serration, which means $\dot\nu$  of this pulsar has 
different values in different epochs. 
The value of $\dot\nu$ increase on MJD 55183 is $\Delta\dot{\nu}_{u}=(191\pm9)\times10^{-15}$\,Hz\,s$^{-1}$, which is 
consistent with the report by \cite{2011ApJS..194...17R}. However, $\dot\nu$ did not experience the 
exponential recovery process like a normal glitch but remained almost steady until MJD 55803, on which $\dot\nu$
suddenly decreased by $\Delta\dot{\nu}_{d}=(-219\pm1)\times10^{-15}$\,Hz\,s$^{-1}$ to the 
value nearly equal to that before MJD 55183. From MJD 56114 to 56398, PSR J1124--5916 behaved similarly,  
with $\Delta\dot{\nu}_{u}=(195\pm6)\times10^{-15}$\,Hz\,s$^{-1}$ 
and $\Delta\dot{\nu}_{d}=(-161\pm10)\times10^{-15}$\,Hz\,s$^{-1}$. 
These sudden changes of $\dot{\nu}$ are very similar to the spin-down rate transition 
of PSR B0540--69 though the fractional amplitudes
($\sim0.4\%$) are about two orders of magnitude smaller \citep{Ge2019}. These variations are 
also similar to PSR B1931+24 \citep{2006Sci...312..549K} 
except that the pulse signal does not disappear. From these similarities, 
we suggest the spin-down state transition of PSR J1124-5916 are probably physically the same as the other 
pulsar state transitions reported in the literatures \citep{2006Sci...312..549K,
2010Sci...329..408L,2015ApJ...807L..27M,2015MNRAS.446.1380P,2016MNRAS.455.1071P,2020ApJ...890...16T}.
We note here that a glitch happened around MJD 58632, whose parameters
are listed in Table \ref{table_timing_para0}, but no further discussions on this glitch will be 
presented in this work.

From Figure \ref{fig0} we know that the state with higher $|\dot{\nu}|$ covers much longer 
time ($\sim80\%$) than the low $|\dot{\nu}|$ state, and the former is thus named normal spin-down state. 
We find that $|\dot{\nu}|$  in the normal spin-down state decreases linearly with time as represented by 
the red line in Figure \ref{fig0} (b). The linear fitting result is listed in Table \ref{table_timing_para0}.
The two states with lower $|\dot{\nu}|$ are called the low spin-down state I 
and the low spin-down state II, respectively. Figure \ref{fig0} also suggests that there should 
be some  sudden variations before MJD 54600. Particularly, the pulsar might switch to 
the low spin-down state around MJD 52180 \citep{2002ApJ...567L..71C}, if the main spin-down state
between MJD 52105--54600 can be also represented by the red line in Figure \ref{fig0} (b). 
The pulsar around MJD 52105 and 53965 might be in normal spin-down 
state considering the uncertainty of the measured parameters \citep{2003ApJ...591L.139H,Abdo2010b}. 

\begin{figure}
\begin{center}
\includegraphics[scale=0.75]{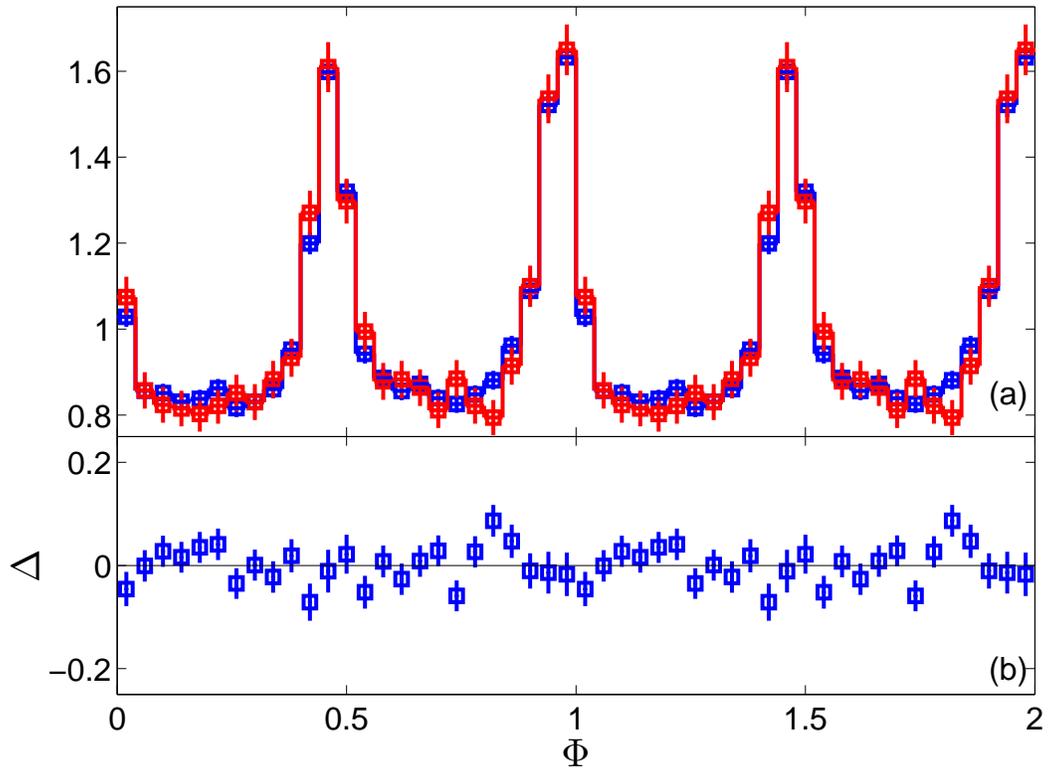}\caption{
The pulse profiles of PSR J1124--5916. Panel (a): The red and blue lines represent the normalized profiles obtained 
from the low spin-down states and the normal spin-down state. The normalized profile is obtained from the preliminary profile divided by its mean value. Panel (b): The differences between two pulse profiles.
\label{fig1}}
\end{center}
\end{figure}

We checked the $\gamma$-ray pulse profiles and fluxes of PSR J1124-5916 in both the normal 
and low spin-down states to see whether there exist any variations. The two pulse profiles 
and their differences are plotted in Figure \ref{fig1}.  
The uniform distribution and the low reduced $\chi^{2}$ value (1.29, for 50 d.o.f.) of the differences 
show that the two profiles are almost identical. The $\gamma$-ray flux 
of PSR J1124-5916 keeps constant whenever in the low spin-down state 
or in the normal spin-down state considering the measurement errors\footnote{https://fermi.gsfc.nasa.gov/ssc/data/access/lat/8yr\_catalog/LcPlot2months\_4FGL\_v21.pdf}.
These features are also similar to PSR B0540--69, whose X-ray 
pulse profile and pulsed flux remain unchanged after state-transition \citep{Ge2019}. 

The long time observations also allow us to measure the braking index of PSR J1124-5916. 
Fitting the results from {\sl Fermi}-LAT, radio and X-ray in MJD 52105 to 58632 by a quadratic polynomial 
gives a braking index of $1.8\pm0.1$ as listed in Table \ref{table_timing_para0}. 
However, because of the state transition activities, this value might be unreliable. Then, the spin evolution between MJD 56398 and 58632 is selected to calculate 
the braking index because the pulsar remains in the normal spin-down state in this time interval, and the resulted braking index is $1.84\pm0.06$ ( Table \ref{table_timing_para0}). Finally, the spin evolution of the whole normal spin-down state is fitted with equation (\ref{eq11}) to calculate the braking index  and the braking index is $1.98\pm0.04$ ( Table \ref{table_timing_para1}), which is consistent with the one obtained between MJD 56398 and 58632 within 2\,$\sigma$. This makes PSR J1124--5916 the thirteenth pulsar with reliable braking 
index \citep{2011ApJ...741L..13E,2017MNRAS.466..147E,Archibald2016,Clark2016}.
Similarly, the braking indices in low spin-down state I and II could be calculated 
from the timing solutions listed in Table \ref{table_timing_para1}. The braking 
index in low spin-down state I is $0.3\pm0.1$, which is smaller than the value in the
normal spin-down state and similar to the value of PSR B0540--69 after phase transition \citep{Ge2019}.
The braking index in low spin-down state II is $-2.7\pm0.9$, 
which is not reliable due to the large timing noise and short 
time range. Under the assumption that the spin-down is due 
to the dipole radiation and pulsar wind production, the variations of the 
braking index imply that the structure of the magnetosphere changes during 
the spin-down state transition \citep{Ge2019,2020MNRAS.tmp..161W}.

\section{Discussions and Summary}

State transitions are very rare events for young pulsars. Besides PSR J1124-5916, 
only PSR B0540-69 is detected experiencing a state transition with $|\dot{\nu}|$ 
increased by $36\%$ in December 2011 within 15\,days \citep{2015ApJ...807L..27M}. 
Although the X-ray luminosity of the pulsar wind surrounding PSR B0540-69 increased gradually 
up to $30\%$ above the mean pre-transition value, the pulsed X-ray flux and profile remain 
unchanged \citep{Ge2019}. The spin-down rate increase of PSR B0540-69 is suggested to be the 
consequence of local changes in the magnetosphere, perhaps in the magnetic pole region, 
which enhance the pulsar wind production and thus the X-ray brightness of the pulsar wind nebula \citep{Ge2019}.  
The {\sl Fermi}-LAT observations show that PSR J1124--5916 has transition 
behaviors similar to PSR B0540--69, such as the non-variation of the pulse profile and flux 
as well as braking index changes. However, both the fractional amplitude of the state 
transition ($~0.4\%$) and the duration of the abnormal spin-down 
state (about 620 and 280 days) of PSR J1124-5916 are much smaller than those of PSR B0540-69, 
which are $36\%$ and $\ge3000$\,days, respectively. 

However, state transition behaviors have been detected in a number of radio pulsars and 
one $\gamma$-ray pulsar \citep{2006Sci...312..549K,2010Sci...329..408L,2020ApJ...890...16T}.
A common feature of these pulsars is that they are quite old, with characteristic age $\gtrsim100$\,kyr.
PSR B1931+24 shows periodical emission behavior, which could switch
off in 10 seconds \citep{2006Sci...312..549K}. On the other hand, some pulsars are reported 
to have gradual $\dot{\nu}$ changes with timescales of about tens of days to 
several years \citep{2010Sci...329..408L,2015MNRAS.446.1380P,2016MNRAS.455.1071P,2020ApJ...890...16T}.
The fractional variation amplitude of $\dot{\nu}$ of PSR J1124--5916 is similar to the amplitudes of some of them.
Considering the common features from these pulsars, the state transition of PSR J1124--5916 could 
be also explained by as the result of changes in the pulsar magnetosphere \citep{1999ApJ...525L.125H,2013ApJ...768..144T}. 
\cite{Ge2019} suggested that, as the radio emission originates probably in the magnetic pole region and the 
high energy emission originates from a broader region such as the outer gap \citep{1986ApJ...300..500C}, 
the no-change of the $\gamma$-ray pulse profile after spin-down state transition does not 
mean that the radio pulse profile will remain steady. Radio monitoring on PSR J1124--5946 will be
helpful to probe the properties of the magnetosphere and the mechanism of spin-down state transition.  

Due to the timing noise and stochastic variations in the spin-down evolution, only twelve 
pulsars have been measured reliable braking indices \citep{2011ApJ...741L..13E,2017MNRAS.466..147E,Archibald2016,Clark2016}.
The braking index of PSR J1124--5916 is less than 3
and close to that of PSR B0540--69 before state transition \citep{2011ApJ...741L..13E}.
This means that the spin evolution of PSR J1124--5916 might be dominated by
a simple magnetic-dipole radiation and pulsar wind braking scenario \citep{2001ApJ...561L..85X,2013ApJ...768..144T}.

In summary, PSR J1124--5916 is the second young
pulsar showing state transition behaviors. The $\dot{\nu}$ evolution of
PSR J1124--5916 shows two states named the normal spin-down state and the low spin-down state. 
The ratio of duration in the normal spin-down state is $\sim80\%$  and in the rest of 
time it switches to the low spin-down state twice. However,  
the $\gamma$-ray pulse profile and flux did not show variations
after state transition within the uncertainty of the measurements. Finally, we measured 
its braking index as $1.98\pm0.04$ utilizing the 
observation of whole normal spin-down state.


\acknowledgments{We thank F. F. Kou for helpful discussions on 
the braking index of pulsars. This work is supported by the
National Key R\&D Program of China (2016YFA0400800)
and the National Natural Science Foundation of China under
grants U1938109, U1838201, U1838202, 11873080, 11903001, U1838104 and U1838101.
We acknowledge the use of the public data from the {\sl Fermi} data archive.}

\clearpage

\begin{table}
\caption{Timing parameters of PSR J1124--5916.}
\scriptsize{} \label{table_timing_para0}
\begin{center}
\begin{tabular}{l l l l l l l}
\hline\hline
Parameters                                        & Value \\
\hline
R.A. & 11:24:39.0 \\
Decl. & -59:16:19 \\
Epoch(MJD)                                      &     55000         \\
Time range                                    &      52105--58632     \\
$\nu$(Hz)                                         &     7.3802155(4) \\
$\dot\nu$($10^{-11}$\,Hz\,s$^{-1}$)&    -4.0978(3)     \\
$\ddot{\nu}$($10^{-22}$\,Hz\,s$^{-2}$)&   4.16(24)     \\
$ n $ &   1.8(1)      \\
\hline

Time range                                    &      56398--58632     \\
Epoch(MJD)                                      &     57500         \\
$\nu$(Hz)                                         &     7.37137457(3) \\
$\dot\nu$($10^{-11}$\,Hz\,s$^{-1}$)&    -4.09186(3)     \\
$\ddot{\nu}$($10^{-22}$\,Hz\,s$^{-2}$)&   4.18(12)     \\
$ n $ &   1.84(6)      \\
\hline
\hline
Glitch & \\
\hline
Time range                                   &         58632--58999\\
Epoch(MJD)                                      &  58632(10)    \\
$\Delta\nu$ ($10^{-6}$\,Hz) &    0.18(17) \\
$\Delta\nu/\nu$ ($10^{-9}$) &    25(21) \\
$\Delta\dot\nu$($10^{-15}$\,Hz\,s$^{-1}$)&  -3.1(5)  \\
$\Delta\dot\nu/\dot{\nu}$($10^{-3}$)&    0.77(13)  \\
$\Delta\nu_{\rm d}$($10^{-6}$\,Hz)&  2.1(2) \\
$\tau$(day)&  87(7) \\
\hline
\end{tabular}
\end{center}
The confidence interval is 68.3\%.
\end{table}

\begin{table}
\caption{The state transition parameters of PSR J1124--5916.}
\scriptsize{} \label{table_timing_para1}
\begin{center}
\begin{tabular}{l l l l l l l}
\hline\hline
Parameters                                        & Value \\
\hline
Normal spin-down state & \\
\hline
Epoch(MJD)                                      &  56600    \\
Time range                                    &         54600--55183, 55803--56114, 56398--58632    \\
$\nu$($10^{-6}$\,Hz)&     7.37455773(3)  \\
$\dot\nu$($10^{-11}$\,Hz\,s$^{-1}$)&    -4.09537(8) \\
$\ddot{\nu}$($10^{-22}$\,Hz\,s$^{-2}$)&   4.50(9)      \\
$\Delta{\nu_{\rm J1}}$($10^{-6}$\,Hz)&   -11.2(2)      \\
$\Delta{\nu_{\rm J2}}$($10^{-6}$\,Hz)&   4.3(1)      \\
$n$&   1.98(4)     \\
\hline
\hline
Low spin-down state I & \\
\hline
Time range                                   &         55183(5)-55803(5)    \\
Epoch(MJD)                                      &  55600    \\
$\nu$(Hz)                                         &     7.3780914669(9) \\
$\dot\nu$($10^{-11}$\,Hz\,s$^{-1}$)&    -4.08593(1)     \\
$\ddot{\nu}$($10^{-22}$\,Hz\,s$^{-2}$)&   0.76(21)     \\
$ n $ &   0.3(1)       \\
$\Delta\nu_{\rm u}$($10^{-9}$\,Hz)&    -27(10)  \\
$\Delta\dot\nu_{\rm u}$($10^{-15}$\,Hz\,s$^{-1}$)&    191(9)  \\
$\Delta\nu_{\rm d}$($10^{-9}$\,Hz)&    -252(8)  \\
$\Delta\dot\nu_{\rm d}$($10^{-15}$\,Hz\,s$^{-1}$)&    -219(1) \\
\hline
\hline
Low spin-down state II & \\
\hline
Time range                                   &         56114(5)-56398(11)    \\
Epoch(MJD)                                      &  56250    \\
$\nu$(Hz)                                         &     7.375794274(4) \\
$\dot\nu$($10^{-11}$\,Hz\,s$^{-1}$)&    -4.07937(4)     \\
$\ddot{\nu}$($10^{-22}$\,Hz\,s$^{-2}$)&  -6(2)     \\
$ n $ &   -2.7(9)       \\
$\Delta\nu_{\rm u}$($10^{-9}$\,Hz)&    363(9)  \\
$\Delta\dot\nu_{\rm u}$($10^{-15}$\,Hz\,s$^{-1}$)&    195(6)  \\
$\Delta\nu_{\rm d}$($10^{-9}$\,Hz)&    1(11)  \\
$\Delta\dot\nu_{\rm d}$($10^{-15}$\,Hz\,s$^{-1}$)&    -161(10)  \\
\hline
\end{tabular}
\end{center}
The confidence interval is 68.3\%.
\end{table}

\end{document}